\documentstyle[12pt,epsf]{article}
 
\def\eps{\epsilon^*}

\begin{document}
\begin{titlepage}
\begin{flushright}
\begin{tabular}{l}
FERMILAB--CONF--98/098--T\\
hep-ph/9803501
\end{tabular}
\end{flushright}
\vskip0.5cm
\begin{center}
  {\Large \bf
                B DECAYS INTO LIGHT MESONS
  }
 
\vspace{3cm}
{\sc Patricia~Ball}\footnote{Address after 1 May 1998: CERN/TH, 1211
Gen\`{e}ve 23, Switzerland}
\\[0.3cm]
\vspace*{0.1cm} {\it
Fermi National Accelerator Laboratory,
  P.O.\ Box 500, Batavia, IL 60510, USA} \\[2cm]
 
\vspace*{3cm}

{\large\bf Abstract:\\[10pt]} \parbox[t]{\textwidth}{
I calculate the form factors describing semileptonic and penguin
  induced decays of $B$ mesons into light pseudoscalar and vector mesons. The
  form factors are calculated from QCD sum rules on the light-cone
  including contributions  up to twist~4,
  radiative corrections to the leading twist contribution and SU(3)
  breaking effects. The theoretical uncertainty is estimated
  to be $\sim\,$(15--20)\%.
}
  \vfill
{\em To be published in the proceedings of XXXI.\ Rencontres
de Moriond: QCD and high energy hadronic interactions, Les Arcs,
France, 21--28 Mar 1998.}
\end{center}
\end{titlepage}

\newpage


Decays of $B$ mesons into light mesons offer
the possibility to access the less well known entries in the CKM quark
mixing matrix like $V_{ub}$ and $V_{ts}$. The measurement of rare
penguin induced $B$ decays may also give hints at new physics in
the form of loop-induced effects. With new data of hitherto unknown
precision from the new experimental facilities
BaBar at SLAC and Belle at KEK expected to be available in the near
future, the demands at the accuracy of theoretical predictions are
ever increasing. The central problem of all such
predictions, our failure to solve nonperturbative QCD,
is well known and so far prevents a rigorous calculation of form
factors from first principles. Theorists thus
concentrate on providing various approximations. The
maybe most prominent of these, simulations of QCD on the lattice,
have experienced considerable
progress over recent years; the current status for $B$ decays is summarized in
\cite{flynn}. It seems, however, unlikely that lattice calculations will
soon overcome their main restriction in describing $b\to u$ and $b\to s$
transitions, namely the effective upper cut-off that  the finite
lattice size imposes on the momentum of the final state meson. The
cut-off restricts lattice predictions of $B$ decay form factors  to rather
large momentum transfer $q^2$ of about 15$\,$GeV$^2$ or larger. The physical
range in $B$ decays, however, extends from 0 to about
20$\,$GeV$^2$, depending on the process; for radiative decays like
$B\to K^*\gamma$ it is exactly 0$\,$GeV$^2$. Still, one may hope to
extract from the lattice data some information on  form factors in
the full physical range, as
their behaviour at large $q^2$ restricts the shape
at small $q^2$ via the analytical properties of a properly chosen
vacuum correlation function. The latter function, however,
also contains  poles and multi-particle cuts whose exact behaviour
is not known, which limits the accuracy of bounds obtained from such
unitarity constraints and until now has restricted their application
to $B\to\pi$ transitions \cite{Laurent,constraints}. The
most optimistic overall theoretical uncertainty one may hope to obtain
from this method is
the one induced by the input lattice results at large $q^2$, which to
date is around 30\% \cite{latticeBpi,Laurent}. A more
model-dependent extension of the lattice
form factors into the low $q^2$ region is discussed in
\cite{latticeparametrizations}.
 
An alternative approach to heavy-to-light transitions
is offered by QCD sum rules on the
light-cone. In contrast and complementary to lattice simulations, it is
just the fact that the final state meson {\em does} have large energy
and momentum of order
$\sim m_B/2$ in a large portion of phase-space that is  used as
starting point (which restricts the method to not too large momentum
transfer, to be quantified below).
The key idea is to consider $b\to u$ and $b\to s$
transitions as hard exclusive QCD processes and to combine the
well-developed description of such processes in terms of perturbative
amplitudes and nonperturbative hadronic distribution amplitudes
\cite{exclusive} (see also \cite{sterman} for a nice
introduction) with the method of QCD sum rules \cite{SVZ} to
describe the decaying hadron. The idea of such ``light-cone sum
rules'' was first
formulated and carried out in \cite{BBK} in a different context
for the process $\Sigma\to
p\gamma$, its first application to $B$ decays was given in
\cite{chernB}. Subsequently, light-cone sum rules were considered for
many $B$ decay processes, see \cite{VMBreview,KRreview} for
reviews.\footnote{There also
exists an extended literature on a more ``direct'' extension
of QCD sum rules to heavy-to-light transitions, which is based on
three-point correlation functions, see e.g.\ \cite{3pt}. The conceptional
restrictions of
these sum rules are discussed in Ref.~\cite{rhoFFs}. They fail to give a
viable description of form factors at small and moderate momentum
transfer.} As light-cone sum rules are based on the light-cone
expansion of a correlation function, they can be systematically
improved by including higher twist contributions and radiative
corrections to perturbative amplitudes. The first calculations in
\cite{chernB,rhoFFs} were done at tree-level and to leading twist~2
accuracy. In \cite{BKR,rest}, twist 3 and 4 contributions to $B\to\pi$
were included,
 in \cite{radcorr}, one-loop radiative corrections to the twist~2
contribution to the form factor $f_+^\pi$ were calculated, and in
\cite{b98}, twist 3 and 4 contributions and next-to-leading
corrections to all $B\to\,$pseudoscalar form factors were
calculated. In these proceedings, I present the results of
Refs.~\cite{b98,survey} for $B\to\,$pseudoscalar and  $B\to\,$vector
transitions, which rely on recent results for twist 3 and 4 vector
meson distribution amplitudes \cite{BBKT,T4}.

Let me begin by defining the form factors.
Let $P$ be a light pseudoscalar meson, i.e.\ $\pi$ or $K$, and $V$ be
a vector meson, i.e.\ $\rho$, $\omega$, $K^*$ or $\phi$; $V_\mu$ and
$A_\mu$ are the appropriate vector and axialvector currents, respectively.
Semileptonic form factors are defined by ($q=p_B-p$)
\begin{eqnarray}
\langle P(p) | V_\mu | B(p_B)\rangle & = & f_+^P(q^2) \left\{
(p_B+p)_\mu - \frac{m_B^2-m_P^2}{q^2} \, q_\mu \right\} +
\frac{m_B^2-m_P^2}{q^2} \, f_0^P(q^2)\, q_\mu,\\
{\rm with\ } f_+^P(0) & = & f_0^P(0),\nonumber\\
\langle V(p) | (V-A)_\mu | B(p_B)\rangle & = & -i \eps_\mu (m_B+m_V)
A_1^V(q^2) + i (p_B+p)_\mu (\eps p_B)\,
\frac{A_2^V(q^2)}{m_B+m_V}\nonumber\\
& + & i
q_\mu (\eps p_B) \,\frac{2m_V}{q^2}\,
\left(A_3^V(q^2)-A_0^V(q^2)\right) +
\epsilon_{\mu\nu\rho\sigma}\epsilon^{*\nu} p_B^\rho p^\sigma\,
\frac{2V^V(q^2)}{m_B+m_V}\\
{\rm with\ }A_3^V(q^2) & = & \frac{m_B+m_V}{2m_V}\, A_1^V(q^2) -
\frac{m_B-m_V}{2m_V}\, A_2^V(q^2),\qquad
A_0^V(0) \ = \ A_3^V(0). \label{eq:A30}
\end{eqnarray}
 The penguin form factors are defined as
\begin{eqnarray}
\langle K(p) | \bar s \sigma_{\mu\nu} q^\nu (1+\gamma_5) b | B(p_B)\rangle
& \equiv &  \langle K(p) | \bar s \sigma_{\mu\nu} q^\nu b |
B(p_B)\rangle\nonumber\\
& = & i\left\{ (p_B+p)_\mu q^2 - q_\mu (m_B^2-m_K^2)\right\} \,
  \frac{f_T^K(q^2)}{m_B+m_K} \label{eq:fT}\\
\langle K^* | \bar s \sigma_{\mu\nu} q^\nu (1+\gamma_5) b |
B(p_B)\rangle & = & i\epsilon_{\mu\nu\rho\sigma} \epsilon^{*\nu}
p_B^\rho p^\sigma \, 2 T_1(q^2)\nonumber\\
& & {} + T_2(q^2) \left\{ \eps_\mu
  (m_B^2-m_{K^*}^2) - (\eps p_B) \,(p_B+p)_\mu \right\}\nonumber\\
& & {} + T_3(q^2)
(\eps p_B) \left\{ q_\mu - \frac{q^2}{m_B^2-m_{K^*}^2}\, (p_B+p)_\mu
\right\}\label{eq:T}\\
{\rm with\ } T_1(0) & = & T_2(0). \label{eq:T1T2}
\end{eqnarray}
The physical range in $q^2$ is $0\leq q^2\leq (m_B-m_{P,V})^2$. Although
there are of course no semileptonic decays $B\to K e \nu$, the above
form factors contribute to e.g.\ $B\to K\ell\bar\ell$.
Recalling the results of perturbative QCD for the $\pi$
electromagnetic form factor as summarized in \cite{sterman},
one may suppose that the
dominant contribution to the above form factors be the exchange of a
hard perturbative gluon between e.g.\ the $u$ quark and the antiquark,
which possibility was advocated for instance in \cite{szcz}. This is,
however, not the case, and it was pointed out already in
Ref.~\cite{chernB} that the dominant contribution
comes from the
so-called Feynman mechanism, where the quark created in the weak
decay carries nearly all of the final state meson's momentum, while
all other quarks are soft, and which bears no perturbative suppression
by factors $\alpha_s/\pi$. In an expansion in the inverse $b$ quark
mass, the contribution from the Feynman mechanism is of the same order
as the gluon exchange contribution  with momentum fraction
of the quark of order $1-1/m_b$, but it
dies off in the strict limit $m_b\to\infty$ due to Sudakov effects.
This means that --- unlike in the case of the electromagnetic $\pi$
form factor --- knowledge of the hadron distribution amplitudes
$$
\phi(u,\mu^2)\sim \int_0^{\mu^2}\!\! dk^2_\perp\, \Psi(u,k_\perp),
$$
where $\Psi$ is the full Fock-state wave function of the $B$ and
$\pi(K)$, respectively, $u$ is the longitudinal momentum fraction
carried by the ($b$ or $u(s)$) quark, $k_\perp$ is the transverse quark
momentum, is not sufficient to calculate the form factors in the form
of overlap integrals
$$
F \sim \int_0^1 du\,dv\, \phi^*_{\pi(K)}(u)\, T_{\rm hard}(u,v; q^2)\,
\phi_B(v)
$$
(with $T_{\rm hard}\propto
\alpha_s$).\footnote{Note also that not much is known
about $\phi_B$, whereas the analysis of light meson distribution
amplitudes is facilitated by the fact that it can be organized in an
expansion in conformal spin, much like the partial wave expansion of
scattering amplitudes in quantum mechanics in rotational spin.}
Instead, in the method of light-cone sum rules, only the light meson
is described by distribution amplitudes.
Logarithms in $k_\perp$ are taken into account by the
evolution of the distribution amplitudes under changes in scale,
powers in $k_\perp$ are taken into account by higher twist
distribution amplitudes. The
$B$ meson, on the other hand, is described like in QCD sum rules
by the pseudoscalar current
$\bar d i\gamma_5 b$ in the unphysical region with virtuality
$p_B^2-m_b^2\sim O(m_b)$, where it can be
treated perturbatively. The real $B$ meson, residing on the physical
cut at $p_B^2=m_B^2$, is then traced by analytical continuation, supplemented
by the standard QCD sum rule tools to enhance its contribution with
respect to that of higher single- or multi-particle states coupling to
the same current.
 
The starting point for the calculation of the form factors
  are thus the
correlation functions ($j_B = \bar d i\gamma_5 b$):
\begin{eqnarray}
{\rm CF}_V & = &
i\int d^4y e^{iqy} \langle P(p)|T[\bar q\gamma_\mu b](y)
j_B^\dagger(0)|0\rangle\ =\
\Pi_+^P (q+2p)_\mu + \Pi_-^P q_\mu,\\
{\rm CF}_T & = &
i\int d^4y e^{iqy} \langle P(p)|T [\bar q\sigma_{\mu\nu} q^\nu b](y)
 j_B^\dagger(x)|0\rangle\ =\ 2 i F_T^P (p_\mu q^2 - (pq) q_\mu),
\end{eqnarray}
and similar ones for vector mesons, which are calculated in an expansion around the light-cone
$x^2=0$. The expansion goes in inverse powers of the $b$ quark
virtuality, which, in order for the light-cone expansion to be
applicable, must be of order $m_b$. This restricts the accessible
range in $q^2$ to $m_b^2-q^2 \stackrel{<}{\sim} O(m_b)$
parametrically. For physical $B$ mesons, I choose $m_b^2-q^2\leq
18\,$GeV$^2$. The technical details of the calculation are described
in \cite{survey}. Important is that in \cite{b98,survey} for the first
time twist 3 and 4 contributions and radiative corrections to the
twist 2 contributions were included. The impact of these corrections
is small: both radiative corrections and twist 4 contributions are at
the 5\% level, which shows that both the light-cone and the
perturbative expansion are under control.

\begin{figure}
\centerline{\epsffile{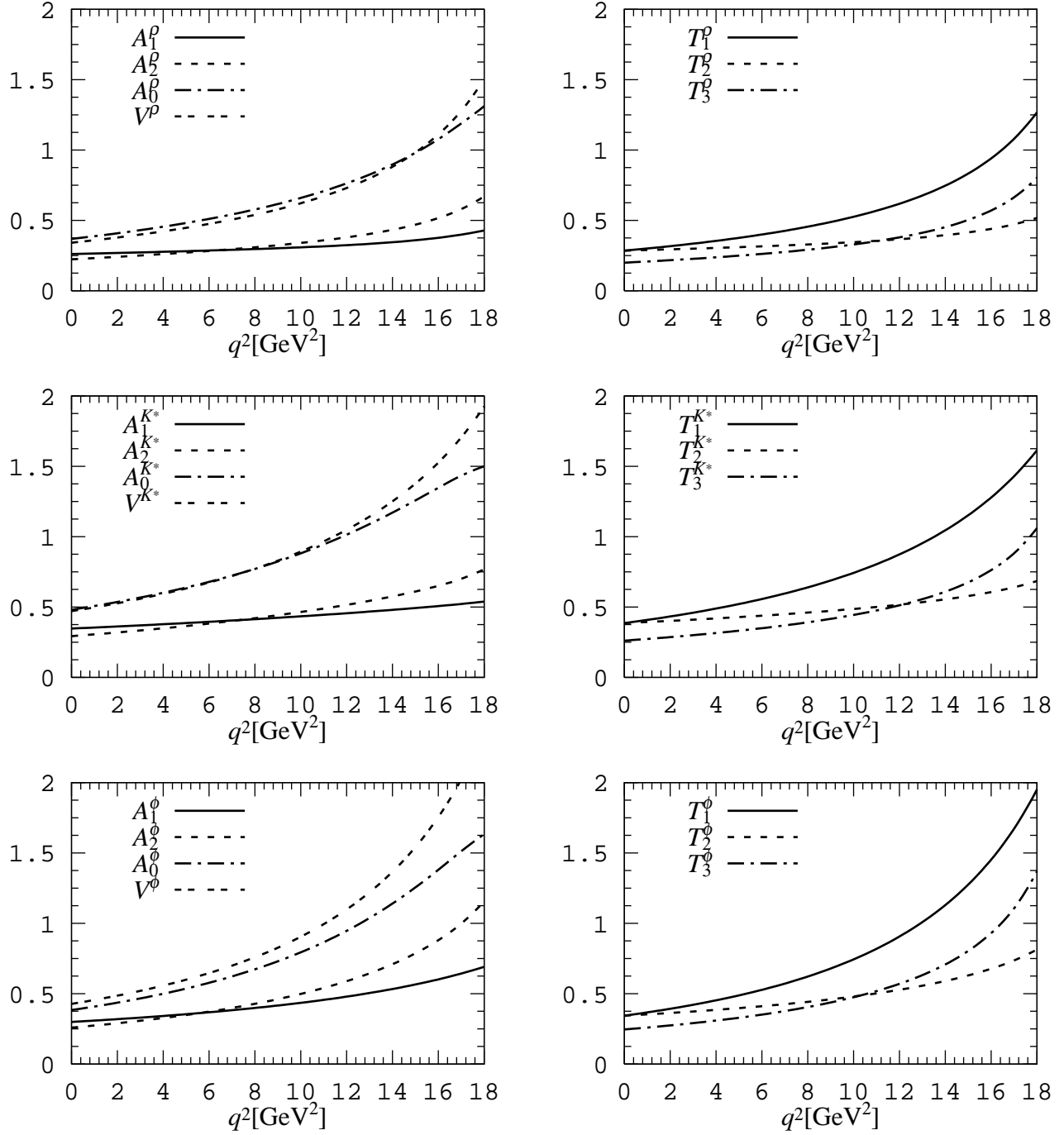}}
\caption{Light-cone sum rule results for $B\to\,$vector meson form
factors. Renormalization scale for $T_i$ is $\mu = m_b =
4.7\,$GeV. Further parameters: $m_b = 4.7\,$GeV, $s_0 = 35\,$GeV$^2$,
$M^2 = 7\,$GeV$^2$.}
\end{figure}

\begin{table}
\caption{Form factors in a three parameter fit. Renormalization scale
for $T_i$ is $\mu = m_b = 4.7\,$GeV.}
\vspace{0.4cm}
$$
\begin{array}{|l|ccc|ccc|l|}
\hline
& F(0) & a_F & b_F & F(0) & a_F & b_F & \\ \hline
f_+^\pi & 0.30\pm 0.04 & 1.35 & \phantom{-}0.27 &
 0.35\pm 0.05 & 1.37 & \phantom{-}0.35 & f_+^K \\
f_0^\pi & 0.30\pm 0.04 & 0.39 & \phantom{-}0.62 &
 0.35\pm 0.05 & 0.40 & \phantom{-}0.41 & f_0^K \\
f_T^\pi & 0.30\pm 0.04 & 1.34 & \phantom{-}0.26 &
 0.39\pm 0.05 & 1.37 & \phantom{-}0.37 & f_T^K \\
\hline
A_1^{\rho} & 0.27 & 0.11 & -0.75 & 0.35 & 0.54 & -0.02 & A_1^{K^*} \\
A_2^{\rho} & 0.23 & 0.77 & -0.40 & 0.30 & 1.02 & \phantom{-}0.08& A_2^{K^*}\\
A_0^\rho & 0.37 & 1.42 & \phantom{-}0.50 & 0.47 & 1.64 & 
\phantom{-}0.94 & A_0^{K^*} \\
V^\rho & 0.34 & 1.32 & \phantom{-}0.19 
 & 0.47 & 1.50 & \phantom{-}0.51& V^{K^*} \\ \hline
T_1^\rho & 0.29 & 1.36 & \phantom{-}0.24 
 & 0.39 & 1.53 & \phantom{-}1.77 & T_1^{K^*}\\
T_2^\rho & 0.29 & 0.08 & -0.94 
 & 0.39 & 0.36 & -0.49 & T_2^{K^*}\\
T_3^\rho & 0.20 & 0.96 & -0.31 
 & 0.26 & 1.07 & -0.16 & T_3^{K^*}\\
\hline
\end{array}
$$
\end{table}

In Fig.~1 I show the form factors for several $B\to\,$vectormeson
transitions as functions of $q^2$ with input parameters as stated in
the caption. SU(3) breaking effects are included by different hadron
distribution amplitudes and amount up to $\sim 10$\%. The remaining
theoretical uncertainty of these form factors is mainly systematic and
dominated by the error introduced by isolating the ground state $B$
meson contribution. This error is estimated to be $\sim 10$\%, and
together with the other uncertainties introduced by the choice of
$m_b$, the QCD sum rule parameters and the hadronic distribution
amplitudes \cite{CZreport,BF,rhoWF,BBKT,T4}, I arrive at a $\sim 20$\%
uncertainty.

The form factors as depicted in the figure lend themselves to a
convenient parametrization in terms of three parameters:
\begin{equation}
F(q^2) = \frac{F(0)}{1-a_F\,\frac{q^2}{m_B^2} +
b_F\left(\frac{q^2}{m_B^2} \right)^2}.
\end{equation}
The corresponding parameters for specific form factors are tabulated
in Tab.~1. The parametrization is acurate to within 1\% for $0\leq
q^2\leq 18\,$GeV$^2$.

Further improvement and refinement of the above results within the
method of light-cone sum rules is difficult and requires in particular
better control over distribution amplitudes. I thus conclude with the
request at the lattice community to feel challenged by the uncertainty
of the old results \cite{latticeold} and to improve them by making
full use if the refined computational implementation of lattice QCD
that has been developed in recent years. 

\section*{Acknowledgments}
Fermilab is
operated by Universities Research Association,
Inc., under contract no.\ DE--AC02--76CH03000
with the U.S.\ Department of Energy.


\section*{References}
\begin{thebibliography}{99}

\bibitem{flynn} J.M.\ Flynn, Talk given at  7th International
Symposium on Heavy Flavor Physics, Santa Barbara, CA, 7-11
Jul 1997, Preprint SHEP-97-25 (hep--lat/9710080);\\
H. Wittig, Lectures given at International
School of Physics, 
Varenna, Italy, 8-18 Jul 1997, Preprint OUTP-97-59-P
(hep--lat/9710088).

\bibitem{Laurent}
L. Lellouch, Nucl.\ Phys.\ {\bf B479} (1996) 353.
 
\bibitem{constraints}
C.G.\ Boyd and M.J.\ Savage, Phys.\ Rev.\ D {\bf 56} (1997) 303;\\
C.G.\ Boyd, B. Grinstein and R.F.\ Lebed, Phys.\ Rev.\ D {\bf 56} (1997)
6895;\\
W.W.\ Buck and R.F.\ Lebed, Preprint hep--ph/9802369.
 
\bibitem{latticeBpi}
D.R.\ Burford et al.\ (UKQCD coll.), Nucl.\ Phys.\ {\bf B447} (1995) 425.
 
\bibitem{latticeparametrizations}
L.  Del Debbio et al.\ (UKQCD coll.), Phys.\ Lett.\ B {\bf 416}
(1998) 392.
 
\bibitem{exclusive} S.J.\ Brodsky and G.P.\ Lepage, in: {\em Perturbative
    Quantum Chromodynamics}, ed.\ by A.H.\ Mueller, p.~93, World
  Scientific (Singapore) 1989;\\
  V.L.\ Chernyak and A.R.\ Zhitnitsky, JETP Lett.\
  {\bf {25}} (1977) 510;
  Yad.\ Fiz.\ {\bf 31} (1980) 1053;\\
  A.V.\ Efremov and A.V.\ Radyushkin, Phys.\ Lett.\ B {\bf 94} (1980)
  245;
  Teor.\ Mat.\ Fiz.~{\bf {42}} (1980) 147;\\
  G.P.\ Lepage and S.J.\ Brodsky, Phys.\ Lett.\ B {\bf 87} (1979) 359;
  Phys.\ Rev.\ D {\bf 22} (1980) 2157;\\
  V.L.\ Chernyak, V.G.\ Serbo and A.R.\ Zhitnitsky, JETP Lett.\ {\bf
    26} (1977) 594; Sov.\ J.\ Nucl.\ Phys.\ {\bf 31} (1980) 552.
 
\bibitem{sterman} G. Sterman and P. Stoler, Preprint ITP-SB-97-49
(hep--ph/9708370).
 
\bibitem{SVZ} M.A.\ Shifman, A.I.\ Vainshtein and V.I.\ Zakharov,
Nucl.\ Phys.\ {\bf B147} (1979) 385; 448; 519.
 
\bibitem{BBK} I.I.\ Balitsky, V.M.\ Braun and A.V.\ Kolesnichenko,
 Nucl.\ Phys.\ {\bf B312} (1989) 509.
 
\bibitem{chernB} V.L.\ Chernyak and I.R.\ Zhitnitsky, Nucl.\ Phys.\
{\bf B345} (1990) 137.
 
\bibitem{VMBreview}
V.M.\ Braun, Preprint NORDITA-98-1-P (hep--ph/9801222).
 
\bibitem{KRreview}
A. Khodjamirian and R. R\"uckl, Preprint
WUE--ITP--97--049 (hep--ph/9801443).
 
\bibitem{3pt}
P. Ball et al., Phys.\ Lett.\ B {\bf 259} (1991) 481;\\
P. Ball, V.M.\ Braun and H.G.\ Dosch, Phys.\ Rev.\ D
{\bf 44} (1991) 3567;\\
P. Ball, V.M.\ Braun and H.G.\ Dosch, Phys.\ Lett.\ B {\bf 273} (1991)
316;\\
P. Ball, Phys.\ Rev.\ D {\bf 48} (1993) 3190;\\
P. Colangelo et al., Phys.\ Rev.\ D {\bf 53} (1996) 3672.
 
\bibitem{rhoFFs} P. Ball and V.M.\ Braun, Phys.\ Rev.\ D {\bf 55}
(1997) 5561.
 
\bibitem{BKR} V.M.\ Belyaev, A. Khodjamirian and R. R\"uckl,
Z.\ Phys.\ C {\bf 60} (1993) 349.
 
\bibitem{rest}
V.M.\ Belyaev et al., Phys.\ Rev.\ D {\bf 51} (1995) 6177;\\
T.M.\ Aliev et al., Phys.\ Lett.\ B {\bf 400} (1997) 194.
 
\bibitem{radcorr}
A. Khodjamirian et al., Phys.\ Lett.\ B {\bf 410} (1997) 275;\\
E. Bagan, P. Ball and V.M.\ Braun, Phys.\ Lett.\ B {\bf 417} (1998) 154.

\bibitem{b98}
P. Ball, Preprint Fermilab--Pub--98/067--T (hep--ph/9802394).

\bibitem{survey} P. Ball and V.M.\ Braun, {\em in preparation}.
 
\bibitem{BBKT} P. Ball et al., Preprint Fermilab--Pub--98/028--T
(hep--ph/9802299).
 
\bibitem{T4} P. Ball, V.M.\ Braun and G. Stoll, {\em in preparation}.
  
\bibitem{szcz}
A. Szczepaniak, E.M.\ Henley and S.J.\ Brodsky, Phys.\ Lett.\ B {\bf
243} (1990) 287.
 
\bibitem{CZreport} V.L.\ Chernyak and A.R.\ Zhitnitsky, Phys.\ Rept.\
  {\bf {112}} (1984) 173.

\bibitem{BF} V.M.\ Braun and I.E.\ Filyanov, Z.\ Phys.\ C {\bf 48}
(1990) 239.

\bibitem{rhoWF} P. Ball and V.M. Braun, Phys.\ Rev.\ D {\bf 54}
(1996) 2182.

\bibitem{latticeold} 
G. Martinelli and C.T.\ Sachrajda, Phys.\ Lett.\ B {\bf 190} (1987)
151;\\
T.A.\ DeGrand and R.D.\ Loft, Phys.\ Rev.\ D {\bf 38} (1988) 954;\\
D. Daniel, R. Gupta and D.G.\ Richards, Phys.\ Rev.\ D {\bf 43} (1991)
3715.


\end{thebibliography}
\end{document}